\documentclass{article}

\usepackage{preprints}
\usepackage{amsmath}
\usepackage[utf8]{inputenc} 
\usepackage[T1]{fontenc}    
\usepackage{hyperref}       
\usepackage{url}            
\usepackage{booktabs}       
\usepackage{amsfonts}       
\usepackage{nicefrac}       
\usepackage{microtype}      
\usepackage{lipsum}
\usepackage{graphicx}
\graphicspath{ {./images/} }
\usepackage{caption}
\usepackage{needspace}
\usepackage[section]{placeins} 
\usepackage{array} 
\usepackage{multirow}

\title{Coordinated Optimization of Departure Sequencing and Section-Track Allocation in Railway Short-Term Concentrated Departure Scenarios Based on QUBO and Hybrid Quantum Algorithms}
\author{
Xiaobin Li$^*$ \\
  School of Transportation Engineering \\
  East China Jiaotong University\\
  Nanchang, Jiangxi 330000, China   \\
  \texttt{yuncifor@outlook.com} \\
\And
 Yanbin Gao\\
  School of Transportation Engineering \\
East China Jiaotong University\\
  Nanchang, Jiangxi 330000, China   \\
  \texttt{specoale296@outlook.com} \\
  \And
 Weiguang Wang\\
  School of Transportation Engineering \\
East China Jiaotong University\\
  Nanchang, Jiangxi 330000, China   \\
  \texttt{Wang0422paper@outlook.com} \\
\And
  Xuechen Liang \\
  School of Transportation Engineering \\
East China Jiaotong University\\
  Nanchang, Jiangxi 330000, China   \\
  \texttt{david.i.ubosi@my.occc.edu} \\
}
\begin{document}
\maketitle
\begin{abstract}
This study examines the coordinated optimization of departure sequencing and section-track allocation in railway short-term concentrated departure scenarios. A quadratic unconstrained binary optimization (QUBO) model is formulated to represent departure-position assignment and section-track selection within a unified binary framework. Because the quality of a dispatching scheme depends on time-dependent operational interactions that cannot be fully captured by a static combinatorial model, a simulation-based evaluation layer is introduced to assess section occupation, intermediate-station waiting, platform-capacity pressure, running-time fluctuations, and delay propagation. Within this layered framework, conventional heuristics, quantum-inspired algorithms, and hybrid algorithms are compared on the same decision structure. The results show that the QUBO model can generate feasible candidate schemes after decoding, while the simulation layer clearly differentiates the operational performance of the competing algorithms under both normal and disturbed conditions. In the tested scenarios, QPSO-QAOA performs best under normal conditions, and the quantum-enhanced methods reduce comprehensive cost by 4.28\%--26.26\% and total delay by 4.37\%--24.25\% on average under dynamic conditions relative to their conventional counterparts. These findings suggest that the integration of QUBO-based modeling and simulation-based evaluation provides a useful methodological framework for railway short-term concentrated departure scheduling, although validation with real operational data remains necessary.The implementation is available at \url{https://github.com/yuncifor/Railway-Short-Term-Based-on-QUBO-and-Hybrid-Quantum-Algorithms}.
\end{abstract}

\section{Introduction}
In railway operations, a short-term concentrated departure scenario arises when multiple trains become ready for departure within a narrow time window and must compete simultaneously for departure priority, section capacity, and downstream station resources. Compared with conventional timetable-planning problems, this setting is more operational, more time-sensitive, and more tightly constrained by local infrastructure. A dispatching decision made at the departure stage can quickly affect section release, intermediate-station occupancy, track conflicts, and subsequent delay propagation. Short-term concentrated departure organization should therefore be regarded not merely as a sequencing problem, but as a coupled dispatching problem involving departure priority, track allocation, and operational feasibility.

Existing studies on railway dispatching, train rescheduling, platforming, and track allocation provide an important foundation for analyzing such problems. Research on train routing and scheduling has shown that railway operational decisions are highly constrained and strongly coupled, especially under saturated and disturbed conditions\textsuperscript{\cite{Cordeau1998TrainRoutingScheduling,Lusby2011RailwayTrackAllocation,Cacchiani2014RailwayReschedulingOverview,Sharma2023PassengerOrientedRescheduling}}. Studies on station routing, platform assignment, and conflict management have further demonstrated that local infrastructure constraints can substantially influence network-level operating performance\textsuperscript{\cite{Zwaneveld2001RoutingTrainsStation,Caprara2011Platforming,DAriano2008ReorderingRerouting,Tornquist2007NTrackedRescheduling}}. In addition, research on real-time rescheduling and robustness has emphasized that dispatching schemes should be assessed not only by nominal efficiency, but also by their ability to mitigate knock-on delays and maintain stability under uncertainty\textsuperscript{\cite{Dewilde2014RobustnessStationAreas,Kroon2008StochasticImprovement,DAriano2007BranchBound,Corman2017DelayManagement}}.

However, most existing studies address timetable adjustment, routing, platforming, or rescheduling as separate subproblems, whereas the short-term concentrated departure problem requires the coordinated treatment of departure sequencing and section-track assignment within a unified decision framework.

From a modeling perspective, the key decisions involved in short-term concentrated departure organization are inherently discrete. Each train must be assigned to exactly one departure position, and each section must be matched with a feasible track or route option subject to resource conflicts and operational penalties. This structure is naturally suited to binary combinatorial optimization. In recent years, quadratic unconstrained binary optimization (QUBO) has attracted increasing attention as a unified modeling framework for encoding discrete decision variables and their pairwise couplings\textsuperscript{\cite{Lucas2014IsingNP,Glover2022QUBOTutorial}}. A major advantage of QUBO is that heterogeneous operational constraints and coordination relationships can be embedded in a single binary objective through penalty terms, thereby allowing sequencing choices, track-selection decisions, and conflict-avoidance requirements to be represented in a mathematically consistent manner. This feature is particularly relevant to railway short-term concentrated departure scheduling. In this context, departure order and section-track allocation cannot be optimized independently: a desirable departure sequence may become infeasible once downstream track usage is considered, whereas a locally feasible track-allocation pattern may lead to excessive waiting or conflict accumulation when combined with an inappropriate departure order. A QUBO-based representation therefore provides a natural way to encode these coupled combinatorial decisions within a unified model. More importantly, it establishes a common solution space in which conventional heuristics, quantum-inspired methods, and hybrid quantum-classical algorithms can be compared on the same decision structure rather than on different reformulations of the problem.

With the development of quantum optimization, QUBO-based methods have gradually been explored in railway and transportation applications. Prior studies have investigated Ising or QUBO formulations for general combinatorial problems\textsuperscript{\cite{Lucas2014IsingNP,Glover2022QUBOTutorial,Kochenberger2014UBQPSurvey}}, railway rescheduling for quantum computing\textsuperscript{\cite{Domino2022RailwayReschedulingQUBO}}, and high-speed train timetable optimization using quantum-oriented solution approaches\textsuperscript{\cite{Xu2023HighSpeedQuantum}}. These studies suggest that quantum algorithms and hybrid approaches may offer promising search mechanisms for large-scale discrete railway decision spaces. Nevertheless, current research remains limited in directly addressing short-term concentrated departure organization, and even fewer studies combine QUBO-based combinatorial modeling with an explicit operational evaluation process to examine whether candidate binary solutions remain effective under time-dependent dispatching dynamics, section-occupation interactions, station-capacity constraints, and delay-propagation effects.

Motivated by this gap, this study develops a layered framework for railway short-term concentrated departure scheduling, in which a QUBO-based decision layer generates candidate combinatorial schemes and a simulation-based evaluation layer examines their operational consequences. This framework addresses a central methodological issue: scheme quality cannot be judged solely by static combinatorial feasibility, because downstream section occupation, intermediate-station waiting, capacity pressure, and delay propagation all depend on time-based operational interactions.

The main contributions of this study are as follows. First, a unified QUBO formulation is established to jointly encode departure sequencing and section-track allocation in short-term concentrated departure scenarios. Second, a simulation-based evaluation mechanism is introduced to assess decoded schemes with respect to temporal feasibility and operational performance beyond the QUBO objective itself. Third, a common comparison framework is provided for conventional heuristics, quantum-inspired methods, and hybrid algorithms under identical modeling and evaluation conditions. Fourth, numerical experiments are conducted under normal, dynamic, robustness, and scale-expansion settings to examine the behavior of different algorithms within the proposed framework. The objective is not to claim the immediate operational deployment of quantum computing in railway dispatching, but rather to evaluate whether QUBO-based modeling, together with quantum-related and hybrid solution strategies, offers a useful methodological direction for this class of scheduling problems.

\section{Preliminaries}
\subsection{Quadratic unconstrained binary optimization}
QUBO is a widely used formulation for discrete combinatorial optimization because it can express binary decision variables and their pairwise couplings within a single quadratic objective. This modeling form has been applied in a variety of optimization settings, including portfolio optimization\textsuperscript{\cite{Lang2022Portfolio,Sakuler2025Portfolio,Aguilera2024Portfolio}}, transportation planning and routing\textsuperscript{\cite{Dixit2023TNDP,Cattelan2024RideHailing,Osaba2024Package}}, and logistics optimization\textsuperscript{\cite{Weinberg2023SupplyChain,MoncayoMartinez2026SupplyChain,NguyenQuang2025AGV}}. In the present study, the main appeal of QUBO lies in its ability to encode multiple coupled dispatching decisions, such as departure-position assignment and section-track selection, within a unified binary framework.\par

The standard QUBO objective is written as
\begin{equation}
\min_{\mathbf{x} \in \{0,1\}} \quad E(\mathbf{x}) = \sum_{i=1}^{n} a_i x_i + \sum_{1 \leq i < j \leq n} b_{ij} x_i x_j
\end{equation}
where \(\mathbf{x} = (x_1, x_2, \ldots, x_n)\) is a binary decision vector with \(x_i \in \{0,1\}\), \(a_i\) denotes the coefficient of the linear term, and \(b_{ij}\) denotes the coefficient of the quadratic interaction term. The objective can be rewritten in matrix form as
\begin{equation}
E(\mathbf{x}) = \mathbf{x}^T Q \mathbf{x}
\end{equation}
where $Q$ is an $n \times n$ symmetric matrix. Although QUBO is unconstrained in form, practical constraints can be embedded into the objective by penalty terms:
\begin{equation}
E(\mathbf{x}) = \sum_{c \in \mathrm{Constraints}} E_c(\mathbf{x})
\end{equation}
In this study, QUBO is used as a combinatorial decision layer rather than a full operational model. It generates candidate binary schemes, while dynamic temporal effects are evaluated separately in the simulation layer.

\subsection{Quantum-inspired and hybrid optimization context}
Quantum computing and quantum optimization have attracted growing attention because they offer new perspectives on difficult combinatorial problems\textsuperscript{\cite{Preskill2018NISQ,Arute2019Supremacy,Blekos2024QAOAReview}}. In this study, however, the focus is not on hardware-level implementation. Instead, the aim is to compare quantum-inspired and hybrid optimization strategies within a common QUBO-based decision framework. Accordingly, quantum-related methods are discussed only to the extent necessary to motivate the algorithmic comparison in later sections.\par

Among these methods, the Quantum Approximate Optimization Algorithm (QAOA)\textsuperscript{\cite{Farhi2014QAOA,Blekos2024QAOAReview}} is a representative hybrid quantum-classical variational approach. Its relevance here is methodological rather than hardware-oriented: QAOA provides a parameterized refinement mechanism for binary combinatorial problems and therefore serves as the conceptual basis for the hybrid algorithms considered in this study. The main analytical question is whether QAOA-related hybridization can improve solution quality within the proposed railway scheduling framework when compared with conventional and quantum-inspired baselines.

\section{Construction of Quantum Scheduling Model}
\subsection{Problem Description}
\label{sec:headings}
This study considers a short-term concentrated departure scenario in which multiple trains become ready for departure within the same scheduling horizon and must compete for departure priority and section-track resources. The experimental line consists of consecutive stations and sections, and the dispatcher must determine both a departure sequence and a section-track allocation plan under limited section capacity, departure-track resources, arrival--departure track capacity, and station-operation constraints. In this setting, scheme quality cannot be judged solely by the combinatorial arrangement itself, because the consequences of a departure sequence are transmitted through section release, intermediate-station dwelling, resource occupation, and delay propagation during operation. The problem is therefore formulated as a layered optimization task: the upper QUBO layer generates candidate combinatorial schemes for departure sequencing and track allocation, and the lower simulation layer evaluates whether those schemes remain operationally attractive once temporal interactions are taken into account.

\subsection{Model Assumptions}

To facilitate the modeling and analysis of railway short-term concentrated departure scheduling and to maintain consistency with the program implementation, the following assumptions are adopted:
\begin{itemize}
\item Train acceleration and deceleration are treated as instantaneous, and their effects are incorporated into section running times or station dwell times.
\item Differences in train type, consist length, payload, and traction performance are not explicitly modeled.
\item Vehicle circulation and rolling-stock reuse are ignored; the vehicle resources assigned to each train are treated as independent.
\item At the beginning of the scheduling horizon, all trains have completed their pre-departure preparation. Intermediate-station dwell times follow the given parameters, and post-arrival operations are not tracked.
\item The number of tracks remains unchanged during the scheduling period, and section resources are modeled as whole-track occupation without further subdivision into block sections.
\end{itemize}

\subsection{Symbol Explanation}
The main symbols and variables used in the model are summarized in Table~\ref{tab:symbols}.

  \begin{table}[htbp]
    \centering
    \caption{Main symbols used in the model}
    \label{tab:symbols}
   \begin{tabular}{>{\centering\arraybackslash}m{0.18\linewidth}@{\hspace{2.8em}}>{\centering\arraybackslash}m{0.42\linewidth}}
     \toprule
    \textbf{Symbol} & \textbf{Explanation} \\
    \midrule
    $S$ & Set of stations \\
    $I$  & Set of trains \\
    $L$  & Set of tracks \\
    $\Gamma $  & Set of sections \\
    $P$  & Set of departure positions \\
    $T$ & Time set\\
    $d_{i,s}$ & Delay time of train $i$ at station $s$\\
    $x$ & Binary logical variable taking values 0 or 1\\
    $i,i'$ & Any two different trains\\
    $l,l'$ & Any two different tracks\\
    \bottomrule
\end{tabular}
\end{table}
\subsection{Binary Encoding of Scheduling Variables}
Based on the defined sets and binary decision variables, the scheduling scenario is encoded into a binary representation that serves as the basis for QUBO model construction. To transform the railway short-term concentrated departure scheduling problem into a standard form suitable for quantum-inspired combinatorial optimization, the core scheduling decisions are represented by binary variables. Each variable corresponds to a specific scheduling behavior or state and takes a value in $\{0,1\}$. For each train, the departure-position variable indicates whether the train departs from a given position in the ordered sequence, and each train must be assigned to exactly one departure position. Likewise, each departure position must be occupied by exactly one train.

For each train and section, a track-allocation variable indicates whether the train uses a particular track in that section, and each train must choose exactly one track in each section. To unify departure sequencing and section-track assignment within the same representation, the two groups of variables are arranged in a fixed order to form a unidirectional binary decision vector of the QUBO model: $\boldsymbol{z} = \left[ x_{i,p},\, y_{i,\tau,l} \right]^{\mathrm{T}}$.
Let the number of trains in one direction be $n$, the number of running sections be $m$, and the number of available tracks in each section be $l$. Then the dimension of the unidirectional combinatorial decision vector is $\dim(\boldsymbol{z}) = n^2 + n \times m \times l$.
In the numerical example, one direction contains 5 trains, 4 running sections, and 2 available tracks in each section; that is, $n = 5$, $m = 4$, and $l = 2$. Therefore,
\[
\dim(\boldsymbol{z}) = 5^2 + 5 \times 4 \times 2 = 25 + 40 = 65.
\]
Accordingly, the unidirectional QUBO model contains 25 departure-sequence variables and 40 section-track-selection variables, for a total of 65 binary decision variables.
\subsubsection{QUBO Binary Variable Definition}
Each variable corresponds to a specific scheduling behavior or state. The binary vector includes 25 departure-sequence variables and 40 track-allocation variables, which together constitute the input variable set of the QUBO model. For each train $i$, the departure-position variable $x_{i,p}$ indicates whether train $i$ departs from position $p$, and each train must be assigned to exactly one departure position. For each departure position $p$, exactly one train may occupy that position.
\begin{equation}
      x_{i,p}=
    \begin{cases} 
        1, & \text{if train } i \text{ departs from position } p \\
        0, & \text{otherwise}.
    \end{cases}   
\end{equation}
The track-allocation variable $y_{i,\tau,l}$ indicates that train $i$ uses track $l$ in section $\tau$, and each train must select exactly one track in each section.
\begin{equation}
     y_{i,\tau,l}=
    \begin{cases} 
        1, & \text{if train } i \text{ occupies track } l \text{ in interval } \tau \\
        0, & \text{otherwise}.
    \end{cases}    
\end{equation}
\subsection{Constraint and Penalty Design}
To ensure both combinatorial validity and operational plausibility, the QUBO decision layer includes four components: departure-sequence uniqueness, section-track uniqueness, an adjacent-section track-switching penalty, and a same-track congestion penalty for the departure section.\par
\subsubsection{Unique constraint on departure order}
To ensure the legality of the departure-priority scheme, each train must correspond to exactly one departure position, and each departure position must be assigned to exactly one train. The departure-sequence uniqueness penalty is written as
\begin{equation}
H_{\mathrm{seq}}
=
\lambda_1 \sum_{i \in I}\left(\sum_{p \in P}x_{i,p}-1\right)^2
+
\lambda_1 \sum_{p \in P}\left(\sum_{i \in I}x_{i,p}-1\right)^2
\end{equation}
These constraints ensure that each train has one and only one priority assignment and that each priority position is occupied by one and only one train, so that the decoded result forms a complete and nonredundant departure-priority sequence. Here, $\lambda_1$ is the departure-sequence uniqueness penalty coefficient, set to 200.
\subsubsection{Unique Constraint for Section-Track Selection}
To ensure that track allocation is uniquely determined in each operating section, each train must select exactly one track in each section. The section-track-selection uniqueness penalty is written as
\begin{equation}
H_{\mathrm{track}}
=
\lambda_2 \sum_{i \in I}\sum_{\tau \in \Gamma}
\left(\sum_{l \in L} y_{i,\tau,l}-1\right)^2
\end{equation}
This condition guarantees that each train has a unique track-selection result in every section, thereby giving the combinatorial solution a clear interpretation in terms of section-resource assignment. Here, $\lambda_2$ is the section-track-selection uniqueness penalty coefficient, set to 200.
\subsubsection{Track Switching Penalty Constraint for Adjacent Sections}
Because frequent switching between tracks in adjacent sections increases operational complexity, an adjacent-section track-switching penalty is introduced in the QUBO layer to suppress excessive switching:
\begin{equation}
H_{\mathrm{switch}}
=
\lambda_3 \sum_{i \in I}\sum_{\tau=1}^{|\Gamma|-1}
\sum_{l \in L}\sum_{\substack{l' \in L\\ l' \neq l}} y_{i,\tau,l}\, y_{i,\tau+1,l'}
\end{equation}
This term imposes a cost when the same train selects different tracks in adjacent sections, thereby reflecting the additional organizational burden created by track switching. As the number of track switches increases, the penalty also increases, so the optimization tends to favor schemes with more continuous track assignment. Here, $\lambda_3$ is the adjacent-section track-switching penalty coefficient, set to 2.
\subsubsection{Congestion Penalty Constraint for Same Track in the Departure Section}
Because track allocation in the departure section has an important influence on subsequent operation under concentrated departure conditions, a same-track congestion penalty is introduced in the QUBO layer to suppress the excessive concentration of trains on the same departure-section track:
\begin{equation}
H_{\mathrm{cong}}
=
\lambda_4 \sum_{i \in I}\sum_{\substack{i' \in I\\ i' > i}}
\sum_{l \in L} y_{i,\tau_0,l}\, y_{i',\tau_0,l}
\end{equation}
This term penalizes pairwise combinations of trains choosing the same departure-section track, thereby reflecting the concentration of departure-section resource usage. Here, $\lambda_4$ is the same-track congestion penalty coefficient in the departure section, set to 12.
\subsection{QUBO Matrix Formulation}
The objective function of the unidirectional combinatorial scheduling layer can be uniformly written in the standard QUBO form:
\begin{equation}
E_{\mathrm{QUBO}}(\boldsymbol{z})
=
H_{\mathrm{seq}} + H_{\mathrm{track}} + H_{\mathrm{switch}} + H_{\mathrm{cong}}
=
\boldsymbol{z}^{\mathrm{T}} \boldsymbol{Q} \boldsymbol{z}
\end{equation}
where $\boldsymbol{z}$ is the unidirectional combinatorial decision vector and $\boldsymbol{Q}$ is the corresponding QUBO coefficient matrix. The diagonal elements represent the linear coefficients of the binary variables, while the off-diagonal elements reflect quadratic couplings between variables. Departure-sequence constraints, section-track-selection constraints, adjacent-section track-switching penalties, and departure-section same-track congestion penalties can all be encoded into $\boldsymbol{Q}$ through appropriate weights, thereby yielding a unified unconstrained quadratic optimization expression.
\subsection{Objective Functions and Evaluation Metrics}
To distinguish combinatorial search from operational evaluation, this study separately defines the objective function of the QUBO layer and the comprehensive evaluation function of the simulation layer. The former characterizes the static combinatorial structure of departure sequencing and section-track allocation, whereas the latter evaluates the operational performance of the decoded scheme during departure-pool advancement, section release, and intermediate-station dwelling. The QUBO objective is
\begin{equation}
\min E_{\mathrm{QUBO}}(\boldsymbol{z})
=
\min \left(
H_{\mathrm{seq}} + H_{\mathrm{track}} + H_{\mathrm{switch}} + H_{\mathrm{cong}}
\right)
=
\min \boldsymbol{z}^\mathrm{T} \boldsymbol{Q} \boldsymbol{z}
\end{equation}

The comprehensive cost of the simulation layer consists of four parts: delay cost, track-change cost, section-fluctuation penalty, and platform-capacity penalty. It characterizes scheme quality in terms of delay control, track-change organization, section-running fluctuation, and station stopping-resource occupation. A lower value indicates a more favorable scheme for reducing departure waiting, suppressing delay propagation, and alleviating station resource pressure. The simulation-layer objective is
\begin{equation}
\min E_{\mathrm{sim}} = E_{\mathrm{delay}} + E_{\mathrm{change}} + E_{\mathrm{fluct}} + E_{\mathrm{platform}}
\end{equation}
\subsubsection{Delay Cost}
The delay term is used to minimize the total train delay, including both departure-station delay and intermediate-station dwell delay:
\begin{equation}
    E_{\mathrm{delay}} = \sum_{i \in I} \left( d_{i}^{\mathrm{origin}} + d_{i}^{\mathrm{mid}} \right)
\end{equation}
where $d_{i}^{\mathrm{origin}}$ denotes the departure-station delay of train $i$, and $d_{i}^{\mathrm{mid}}$ denotes its accumulated delay at intermediate stations.

\subsubsection{ Track Switching Cost}
The track-change term is used to reduce the number of track transitions between adjacent sections, thereby reducing scheduling complexity and operational cost:
\begin{equation}
    E_{\mathrm{change}} = \sum_{i \in I} \sum_{\tau=1}^{|\Gamma|-1} \left( 1 - \sum_{l \in L} y_{i,\tau,l} \, y_{i,\tau+1,l} \right)
\end{equation}

\subsubsection{ Section Fluctuation Cost}
Train running times in sections are based on preset values, but the actual running times are not fixed. Trains may appropriately accelerate or decelerate under specific conditions, so the actual running time fluctuates within a reasonable range. The section-fluctuation term is written as
\begin{equation}
    E_{\mathrm{fluct}} = \sum_{i \in I} \sum_{\tau \in \Gamma} \left| t_{i,\tau}^{\mathrm{actual}} - t_{i,\tau}^{\mathrm{base}} \right|
\end{equation}
This term characterizes the influence of running-time fluctuation on the comprehensive evaluation. A larger deviation between actual and baseline running time leads to a larger fluctuation penalty, indicating that the combinatorial scheme is more likely to incur additional cost under disturbed conditions.

\subsubsection{ Platform Line Occupancy Cost}
The platform-capacity constraint limits the number of trains that can dwell simultaneously at an intermediate station. When the number of dwelling trains exceeds the normal threshold, standby platform lines must be activated and the following cost is incurred:
\begin{equation}
E_{\mathrm{platform}} = \sum_{s \in S} \sum_{t \in T} \max\left( 0, k_s(t) - k_0 \right)
\end{equation}
where $k_0$ is the platform-capacity threshold and $k_s(t)$ denotes the number of trains dwelling simultaneously at station $s$ at time $t$.

\section{Model Solution}
\subsection{Solution Idea}
\begin{figure}[!htbp]
    \centering
    \includegraphics[width=1\linewidth]{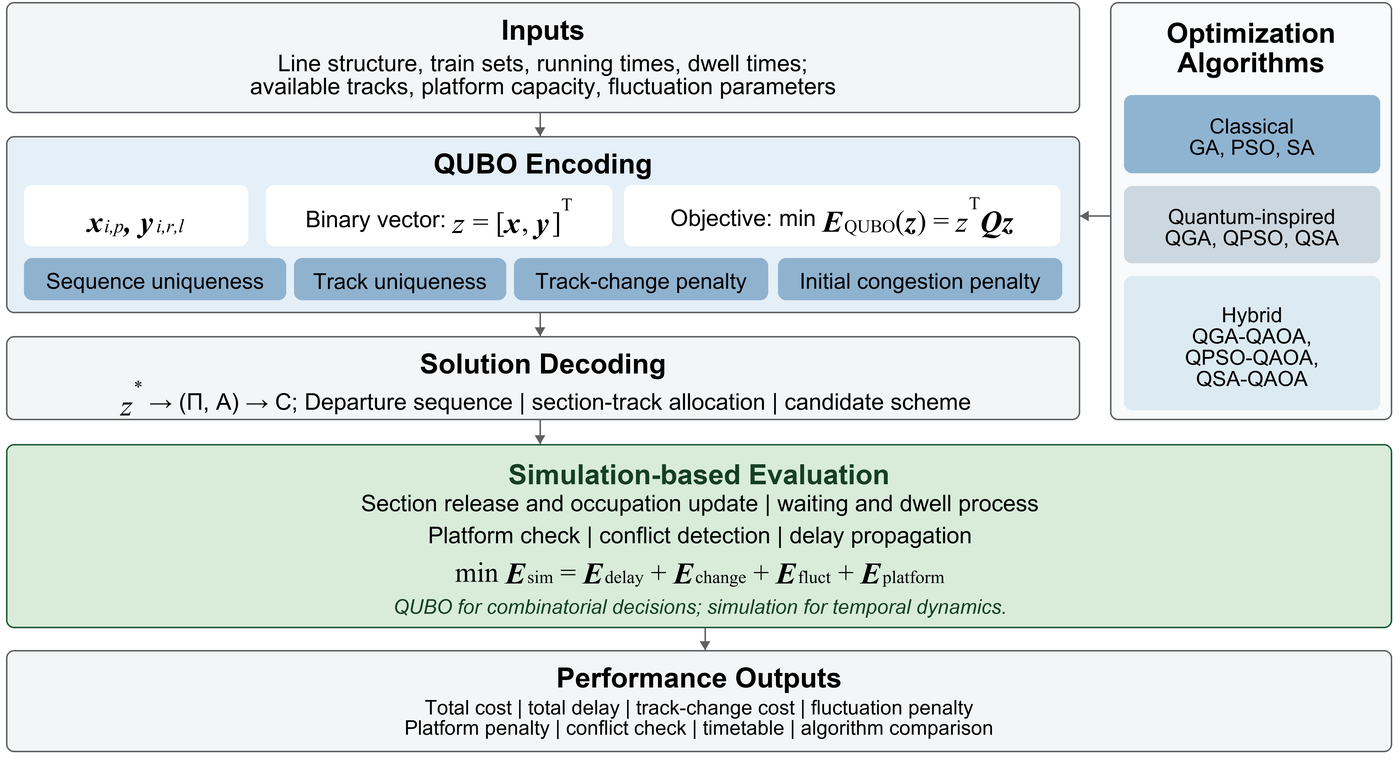}
    \caption{Bilevel framework and solution process for QUBO-based railway short-term concentrated departure scheduling}
    \label{fig:qubo_bilevel_framework}
\end{figure}

Based on the layered framework described above, the solution procedure for the railway short-term concentrated departure problem is implemented in two stages. The aim is not only to obtain a high-quality QUBO combinatorial solution, but also to evaluate the decoded scheme through operational simulation in terms of waiting propagation, resource occupation, and feasibility. The overall layered framework and solution flow are shown in Figure~\ref{fig:qubo_bilevel_framework}.
\FloatBarrier
\subsection{Algorithm Setup}
To compare the applicability of different search mechanisms in the combinatorial decision layer, this study considers a conventional genetic algorithm, conventional particle swarm optimization, conventional simulated annealing, a quantum genetic algorithm, quantum particle swarm optimization, quantum simulated annealing, and several hybrid algorithms under a unified coding framework. All algorithms solve the same QUBO combinatorial problem, and their outputs share the same variable definitions and decoding procedure.\par
Genetic-based methods iteratively update populations through selection, crossover, and mutation, making them suitable for discrete combinatorial search. Particle-swarm-based methods update candidate solutions according to individual-best and global-best information, thereby supporting collaborative population-based search. Simulated-annealing-based methods gradually improve solutions through neighborhood perturbation and probabilistic acceptance and therefore exhibit relatively strong local-search capability. Quantum-inspired methods do not alter the problem formulation itself, but introduce quantum concepts into state representation, update rules, or neighborhood generation to enhance search capability in complex combinatorial spaces. The hybrid algorithms further refine the solutions obtained in the previous stage through QAOA-related local improvement.
\subsection{QUBO Encoding and Decoding}
The QUBO model returns a binary vector rather than a directly usable train operation diagram or timetable. Before entering the simulation layer, this binary solution must be decoded into a combinatorial scheme with practical scheduling meaning. In the unidirectional case, there are 65 binary variables in total, of which the first 25 correspond to departure-sequence variables and the remaining 40 correspond to section-track-selection variables.

For the departure-order part, the index mapping is
\[
k = i \times 5 + p,\quad k \in [0, 24].
\]
If the corresponding entry takes the value 1, train \(i\) is assigned to departure position \(p\). By traversing indices 0 to 24, the departure position of each train can be identified and the priority sequence can be reconstructed accordingly.

For the section-track part, the index mapping is
\[
k = 25 + i \times 8 + \tau \times 2 + l,\quad k \in [25, 64].
\]
If the corresponding entry takes the value 1, train \(i\) selects track \(l\) in section \(\tau\). By traversing indices 25 to 64, the track-selection result of each train in each section can be identified, thereby forming a complete track-allocation scheme. After decoding, the binary vector is transformed into a candidate combinatorial scheme with explicit scheduling meaning.
\subsection{QUBO and Operation Simulation}
The QUBO layer is responsible for the combinatorial optimization of departure sequencing and section-track allocation, whereas the simulation layer converts the combinatorial scheme into an operational process and then evaluates the corresponding cost and feasibility.
The binary solution obtained by the QUBO layer can be decoded into a combinatorial scheme:
\begin{equation}
C = (\Pi, A)
\end{equation}
where $\Pi$ denotes the priority sequence of trains in the departure waiting pool, and $A$ denotes the track allocation scheme of trains on each operation section.\par
In the QUBO layer, the uniqueness constraints and the associated combinatorial penalty terms are encoded directly. Dynamic evaluation items closely related to actual operation, such as section-occupation conflicts and platform-capacity penalties, are not statically embedded in the QUBO matrix because they depend on actual arrival and departure times, section release states, and station-dwelling processes. By contrast, the QUBO layer describes only the two static combinatorial decisions of departure sequence and section-track selection.\par
Based on the priority relation and track-selection result contained in the decoded scheme \(C=(\Pi,A)\), the simulation layer gradually generates the actual train arrival--departure process while taking into account departure-section resource states, section release conditions, intermediate-station dwell rules, and platform-capacity constraints. In the simulation, the beginning of the experimental horizon is uniformly set to \(t=0\), when all trains enter the departure waiting pool. Whether a train can depart depends on the departure-section track resource and its release state. If multiple departure tracks are simultaneously available, concurrent departure is allowed. After a train leaves a section, the release time of the corresponding track is updated, and a following train can enter the section only after that track is released:
\begin{equation}
 t_{i,s}^{\mathrm{dep}} \geq t_{\tau,l}^{\mathrm{release}}   
\end{equation}
where \(t_{i,s}^{\mathrm{dep}}\) denotes the departure time of train \(i\) from station \(s\), and \(t_{\tau,l}^{\mathrm{release}}\) denotes the release time of track \(l\) in section \(\tau\). The time at which a train enters a section is jointly determined by its readiness and the track-release time:
\begin{equation}
 t_{i,s}^{\mathrm{dep}} = \max\left( t_{i,s}^{\mathrm{arr}} + t_{i,s}^{\mathrm{stop}},\; t_{\tau,l}^{\mathrm{release}} \right) + t_{i,s}^{\mathrm{delay}}   
\end{equation}
where \(t_{i,s}^{\mathrm{arr}}\) is the arrival time of train \(i\) at station \(s\), \(t_{i,s}^{\mathrm{stop}}\) is its dwell time, and \(t_{i,s}^{\mathrm{delay}}\) is its delay at station \(s\). After arriving at an intermediate station, a train must first complete its required dwell operation. If the track in the next section has not been released when the dwell process is finished, the train continues waiting in the station, and the corresponding waiting time is counted as intermediate-station delay. In this way, resource shortages in upstream sections can propagate backward through section-release and station-waiting mechanisms.
\subsection{Result Statistics, Feasibility Check and Scheme Output}
After the simulation is completed, the actual arrival--departure process of each train is summarized. The waiting time at the departure station, the waiting time at intermediate stations, the number of track changes between adjacent sections, the section-running-time fluctuation penalty, and the intermediate-station platform-capacity penalty are recorded separately. These records are then aggregated into indicators such as comprehensive cost, total delay, total track-change cost, total fluctuation penalty, and total platform-capacity penalty. Based on the simulated operating results, the program further checks whether conflicts exist in section-track occupation. Let the occupation intervals of train \(i\) and train \(i'\) on the same section and track be \([t_{i,s}^{\mathrm{dep}}, t_{i,s}^{\mathrm{arr}}]\) and \([t_{i',s}^{\mathrm{dep}}, t_{i',s}^{\mathrm{arr}}]\), respectively. If the two intervals overlap,$[t_{i,s}^{\mathrm{dep}}, t_{i,s}^{\mathrm{arr}}] \cap [t_{i',s}^{\mathrm{dep}}, t_{i',s}^{\mathrm{arr}}] \neq \emptyset$
a section conflict is identified. Otherwise, the corresponding section-track occupation arrangement is regarded as feasible. In addition, train operation diagrams, station-level scheduling details, and experimental statistics can be exported from the simulation results to support subsequent algorithm comparison and scheme analysis.
\section{Case Study}
\subsection{Experimental Scenario and Data}
To verify the effectiveness of the proposed model and method, two scenarios are considered: a normal scenario and a dynamic scenario. The case study does not correspond to the original timetable of a specific railway line or station. Instead, it is an abstract test network constructed for methodological validation of the railway short-term concentrated departure scheduling problem. The purpose is to provide a representative scenario in which the applicability of the model and algorithms can be examined.

The experimental line consists of five stations, denoted as A, B, C, D, and E, forming four consecutive running sections between adjacent stations. Five trains are considered in each direction, and two alternative tracks are arranged in each section for the A--E and E--A directions.

Although the structure is moderately simplified, it preserves the core organizational characteristics of the short-term concentrated departure problem, including departure-release competition, continuous section connection, intermediate-station dwell constraints, and bidirectional resource coupling.

The parameter settings are chosen to reflect the main operational tensions in the concentrated departure scenario while facilitating performance comparison among different methods, and they are determined through scenario construction together with pre-experiment calibration. Section running times characterize the basic train-movement process in consecutive sections. Dwell times describe receiving, dispatching, and stopping operations at intermediate stations. Arrival--departure track capacity represents station stopping-resource constraints. The fluctuation rate is used to simulate running-time disturbances. Penalty coefficients balance the influence of track-switching cost, same-track congestion, and operational conflicts on solution quality. The train operation information is listed in Table~\ref{tab:train_operation_information}.
\begin{table}[htbp]
    \centering
\caption{Train operation information}
\label{tab:train_operation_information}
    \begin{tabular}{c c c c}
     \toprule
    Train & Route & Interval travel time/min & Duration of stay at each station/min \\
    \midrule
    21    & A-E   & 5                        & B:2;C:5;D:2                          \\
    22    & A-E   & 5                        & B:3;C:3;D:5                          \\
    23    & A-E   & 5                        & B:2;C:5;D:3                          \\
    24    & A-E   & 5                        & B:3;C:2;D:2                          \\
    25    & A-E   & 5                        & B:4;C:5;D:3                          \\
    31    & E-A   & 5                        & B:2;C:5;D:2                          \\
    32    & E-A   & 5                        & B:3;C:3;D:5                          \\
    33    & E-A   & 5                        & B:2;C:5;D:3                          \\
    34    & E-A   & 5                        & B:3;C:2;D:2                          \\
    35    & E-A   & 5                        & B:4;C:5;D:3                          \\
    \bottomrule
    \end{tabular}
\end{table}
\subsection{Algorithm Parameter Configuration}
To maintain comparability across algorithms, a set of moderate and stable baseline parameters is adopted instead of aggressively tuning any single method for best-case performance. The parameter settings are listed in Table~\ref{tab:algorithm_parameters}.

\begin{table}[htbp]
  \centering
  \caption{Algorithm parameter settings}
  \label{tab:algorithm_parameters}
  \begin{tabular}{ccc}
    \toprule
    \textbf{Algorithm} & \textbf{Parameter} & \textbf{Value} \\
    \midrule
    \multirow{5}{*}{Quantum Genetic Algorithm }
    & Population size & 100 \\
    & Iterations & 100 \\
    & Learning rate & 0.1 \\
    & Elite ratio & 0.05 \\
    & Quantum mutation rate & 0.01 \\
    \cline{2-3}
    \multirow{6}{*}{Quantum Particle Swarm }
    & Number of particles & 100 \\
    & Maximum iterations & 100 \\
    & Initial disturbance scale & 0.5 \\
    & Minimum disturbance scale & 0.01 \\
    & Cognitive coefficient & 0.5 \\
    & Social coefficient & 0.5 \\
    \cline{2-3}
    
    \multirow{6}{*}{Quantum Simulated Annealing }
    & Initial temperature & 100.0 \\
    & Cooling coefficient & 0.99 \\
    & Maximum iterations & 1000 \\
    & Number of superposed states & 5 \\
    & Crossover probability & 0.8 \\
    & Mutation rate & 0.1 \\
    \cline{2-3}
    
    \multirow{4}{*}{Hybrid Algorithm}
    & Local search iterations & 50 \\
    & Number of neighborhood candidates & 10 \\
    & Disturbance probability & 0.1 \\
    & Early stopping patience & 10 \\
    \bottomrule
  \end{tabular}
\end{table}

\subsection{Results and Analysis of the Normal Scenario}
Under normal operating conditions, section running times follow their baseline values and no random fluctuation is introduced. The resulting experiment therefore reflects the relative organizational performance of different combinatorial schemes under deterministic resource constraints. The results are presented in Table~\ref{tab:normal_results}.
\begin{table}[htbp]
  \centering
  \caption{Experimental results under the normal scenario}
  \label{tab:normal_results}
  \begin{tabular}{cccc}
    \toprule
    Algorithm & Total Cost & Solving Time / s & Total Delay / min \\
    \midrule
    GA        & 107 & 0.527685 & 84 \\
    PSO       & 97  & 0.845927 & 79 \\
    SA        & 126 & 0.007769 & 93 \\
    QGA       & 60  & 1.607541 & 56 \\
    QPSO      & 66  & 1.651364 & 64 \\
    QSA       & 72  & 0.304264 & 64 \\
    QGA-QAOA  & 57  & 1.906732 & 55 \\
    QPSO-QAOA & 50  & 1.917093 & 50 \\
    QSA-QAOA  & 83  & 0.290226 & 71 \\
    \bottomrule
  \end{tabular}
\end{table}

The normal-scenario results show that the proposed QUBO model yields solvable combinatorial structures for the tested short-term concentrated departure setting. After decoding, the binary solutions obtained by each algorithm can be converted into departure-order and section-track-allocation schemes with clear scheduling meaning and can then be evaluated in the simulation layer through timetables and operational indicators. This provides evidence that the combinatorial decision layer and the simulation layer are computationally compatible within the tested framework.

After confirming that all methods can produce feasible candidate solutions, the comparison reveals clear differences in solution quality. In terms of total cost and total delay, QPSO-QAOA achieves the best performance in the normal scenario, with a total cost of 50 and a total delay of 50 min. QGA-QAOA and QGA rank next, with total costs of 57 and 60 and total delays of 55 and 56 min, respectively. By contrast, the conventional simulated annealing algorithm yields the highest total cost and total delay among the tested methods.

These results indicate that, under the unified QUBO formulation and simulation-based evaluation setting, the quantum-inspired and hybrid methods achieve better solution quality than the conventional baselines in the tested normal scenario.

Because QPSO-QAOA gives the best overall performance in the normal scenario, it is selected as an illustrative example for detailed timetable analysis. Due to space limitations, only the detailed results of this method are presented here. The upward and downward train timetables are reported in Tables~\ref{tab:upward_schedule} and \ref{tab:downward_schedule}, respectively.

\begingroup
\captionsetup{type=table,hypcap=false}
\centering
\small
\setlength{\tabcolsep}{4pt}
\renewcommand{\arraystretch}{0.92}
\caption{Upward train schedule}
\label{tab:upward_schedule}
\begin{tabular}{lcccccc}
    \toprule
    Train & Station & Track & Arrival Time & Stop/min & Delay/min & Departure Time \\
    \midrule
    \multirow{5}{*}{21}
    & A & 2 & -- & -- & 0 & 0 \\
    & B & 2 & 5 & 2 & 0 & 7 \\
    & C & 2 & 12 & 5 & 0 & 17 \\
    & D & 2 & 22 & 2 & 0 & 24 \\
    & E & - & 29 & - & - & - \\
    \cline{2-7}
    \multirow{5}{*}{22}
    & A & 2 & -- & -- & 5 & 5 \\
    & B & 2 & 10 & 3 & 0 & 13 \\
    & C & 2 & 18 & 3 & 1 & 22 \\
    & D & 2 & 27 & 5 & 0 & 32 \\
    & E & - & 37 & -- & -- & -- \\
    \cline{2-7}
    \multirow{5}{*}{23}
    & A & 1 & -- & -- & 5 & 5 \\
    & B & 1 & 10 & 2 & 1 & 13 \\
    & C & 1 & 18 & 5 & 0 & 23 \\
    & D & 1 & 28 & 3 & 0 & 31 \\
    & E & - & 36 & -- & -- & -- \\
    \cline{2-7}
    \multirow{5}{*}{24}
    & A & 1 & -- & -- & 0 & 0 \\
    & B & 1 & 5 & 3 & 0 & 8 \\
    & C & 1 & 13 & 2 & 0 & 15 \\
    & D & 1 & 20 & 2 & 0 & 22 \\
    & E & - & 27 & -- & -- & -- \\
    \cline{2-7}
    \multirow{5}{*}{25}
    & A & 1 & -- & -- & 10 & 10 \\
    & B & 1 & 15 & 4 & 0 & 19 \\
    & C & 1 & 24 & 5 & 0 & 29 \\
    & D & 1 & 34 & 3 & 0 & 37 \\
    & E & - & 42 & -- & -- & -- \\
  \bottomrule
  \end{tabular}
\par
\endgroup

\vspace{0.5em}
\Needspace{18\baselineskip}
\begingroup
\captionsetup{type=table,hypcap=false}
\centering
\small
\setlength{\tabcolsep}{4pt}
\renewcommand{\arraystretch}{0.92}
\caption{Downward train schedule}
\label{tab:downward_schedule}
\begin{tabular}{lcccccc}
    \toprule
    Train & Station & Track & Arrival Time & Stop/min & Delay/min & Departure Time \\
    \midrule
    \multirow{5}{*}{31}
    & E & 3 & -- & -- & 5 & 5 \\
    & D & 3 & 10 & 2 & 1 & 13 \\
    & C & 3 & 18 & 5 & 0 & 23 \\
    & B & 3 & 28 & 2 & 1 & 31 \\
    & A & - & 36 & -- & -- & -- \\
    \cline{2-7}
    \multirow{5}{*}{32}
    & E & 3 & -- & -- & 0 & 0 \\
    & D & 3 & 5 & 3 & 0 & 8 \\
    & C & 3 & 13 & 3 & 0 & 16 \\
    & B & 3 & 21 & 5 & 0 & 26 \\
    & A & - & 31 & -- & -- & -- \\
    \cline{2-7}
    \multirow{5}{*}{33}
    & E & 3 & -- & -- & 10 & 10 \\
    & D & 3 & 15 & 2 & 1 & 18 \\
    & C & 3 & 23 & 5 & 0 & 28 \\
    & B & 3 & 33 & 3 & 0 & 36 \\
    & A & - & 41 & -- & -- & -- \\
    \cline{2-7}
    \multirow{5}{*}{34}
    & E & 4 & -- & -- & 5 & 5 \\
    & D & 4 & 10 & 3 & 1 & 14 \\
    & C & 4 & 19 & 2 & 3 & 24 \\
    & B & 4 & 29 & 2 & 1 & 32 \\
    & A & - & 37 & -- & -- & -- \\
    \cline{2-7}
    \multirow{5}{*}{35}
    & E & 4 & -- & -- & 0 & 0 \\
    & D & 4 & 5 & 4 & 0 & 9 \\
    & C & 4 & 14 & 5 & 0 & 19 \\
    & B & 4 & 24 & 3 & 0 & 27 \\
    & A & - & 32 & -- & -- & -- \\
  \bottomrule
  \end{tabular}
\par
\endgroup
\textit{Note:} Time values are reported in minutes after 8:00. 
\begin{figure}[!htbp]
    \centering
    \includegraphics[width=1\linewidth]{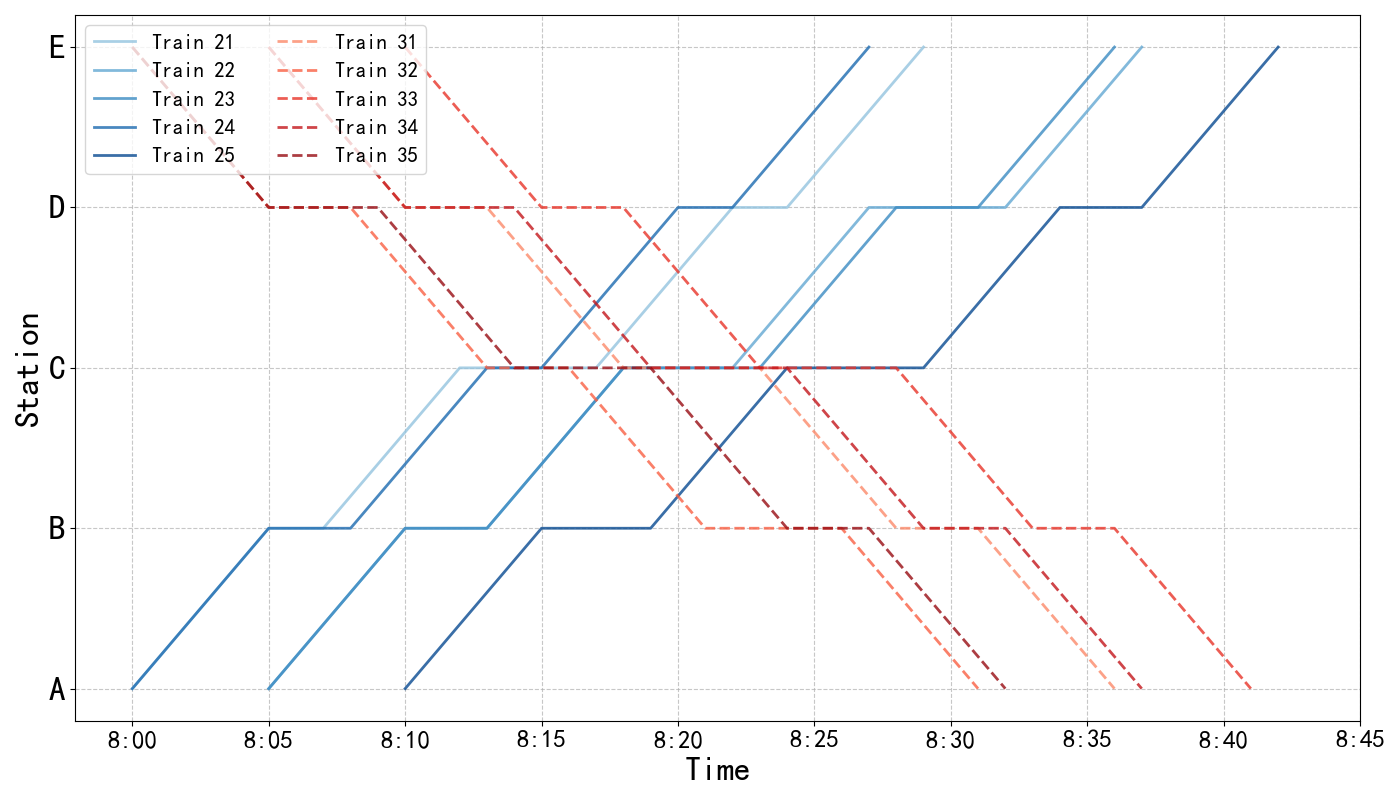}
    \caption{Train operation diagram solved by QPSO-QAOA algorithm}
    \label{fig:Train operation diagram solved by QPSO-QAOA algorithm}
\end{figure}
\FloatBarrier
A departure time of 0 means that the train departs from the station at 8:00, and an arrival time of 28 means that the train arrives at the station at 8:28.
Further analysis of the timetable generated by the QPSO-QAOA algorithm shows that the resulting scheme does not rely on a simple serial release pattern for individual trains. Instead, it implements staggered simultaneous departures whenever resources are available in the initial section. In the upward direction, Trains 21 and 24 depart simultaneously at 8:00, Trains 22 and 23 enter operation at 8:05, and Train 25 departs at 8:10. In the downward direction, Trains 32 and 35 depart concurrently at 8:00, Trains 31 and 34 depart at 8:05, and Train 33 enters operation at 8:10. These results show that the departure-order and track-allocation scheme generated by the QUBO combinatorial layer can be decoded into an executable operating structure while making effective use of multi-track resources in the initial section, thereby avoiding the compression of all trains into a single departure chain.

From the perspective of delay distribution, waiting time is not evenly distributed across all trains, but is concentrated on a limited number of trains and local nodes. In the upward direction, Train 25 has the largest total delay, which mainly originates from waiting at the departure station. Trains 22 and 23 experience only minor local delays at intermediate stations, whereas Trains 21 and 24 maintain more continuous movement across the middle sections. In the downward direction, Trains 33 and 34 experience relatively larger delays, while Trains 32 and 35 remain close to continuous operation. These results indicate that the decoded QUBO solutions are operationally interpretable in the tested scenario and can generate scheduling schemes with clear organizational meaning.
\subsection{Results and Analysis of the Dynamic Scenario}
\begin{figure}[!htbp]
    \centering
    \includegraphics[width=1\linewidth]{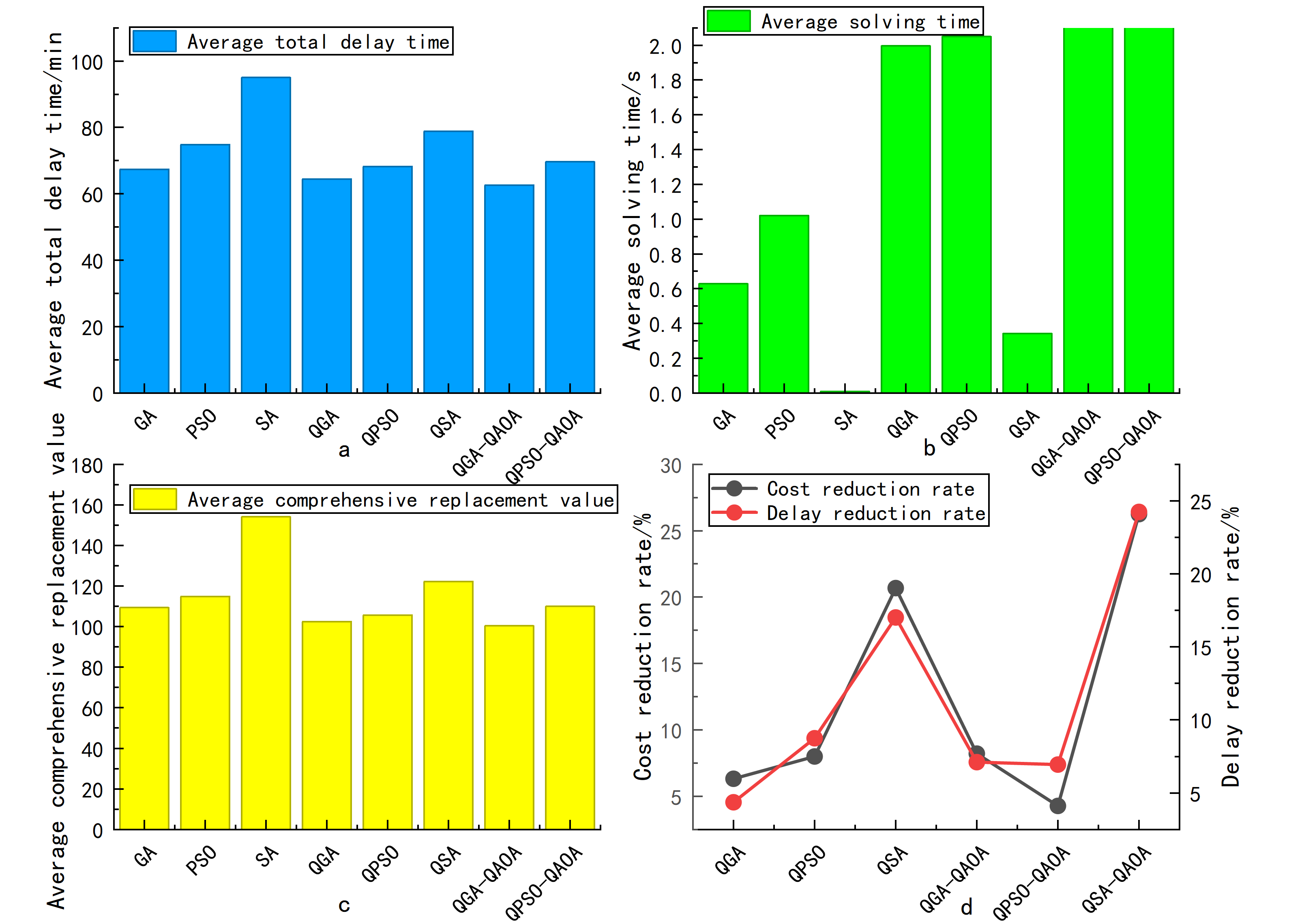}
    \caption{Multi-indicator performance comparison in dynamic experiments}
    \label{fig:Multi-indicator performance comparison in dynamic experiments}
\end{figure}

In the dynamic operation scenario, section running times are allowed to fluctuate by 20\% around their baseline values. Each algorithm is run 100 times, and the average result is reported. The purpose of this experiment is to examine the stability of different algorithms under disturbed conditions and to compare the practical performance of the corresponding scheduling schemes rather than relying on single-run outcomes. Because random fluctuations lead to run-to-run variation, only aggregated statistics are presented. The dynamic experimental results are shown in Table~\ref{tab:dynamic_results} and Fig.~\ref{fig:Multi-indicator performance comparison in dynamic experiments}.

\begin{table}[htbp]
  \centering
  \caption{Dynamic experimental results}
  \label{tab:dynamic_results}
  \resizebox{\linewidth}{!}{%
  \begin{tabular}{lccccc}
    \toprule
    Algorithm & Avg. Total Cost & Cost Reduction/\% & Avg. Solving Time/s & Avg. Delay Time/min & Delay Reduction/\% \\
    \midrule
    GA        & 109.4015  & --    & 0.628099  & 67.401  & -- \\
    PSO       & 114.8825  & --    & 1.021248  & 74.833  & -- \\
    SA        & 154.1340  & --    & 0.009386  & 95.014  & -- \\
    QGA       & 102.4755  & 6.33  & 1.996864  & 64.453  & 4.37 \\
    QPSO      & 105.6825  & 8.01  & 2.050452  & 68.274  & 8.76 \\
    QSA       & 122.2425  & 20.69 & 0.342383  & 78.849  & 17.02 \\
    QGA-QAOA  & 100.4085  & 8.22  & 2.132002  & 62.602  & 7.12 \\
    QPSO-QAOA & 109.9705  & 4.28  & 2.199792  & 69.622  & 6.96 \\
    QSA-QAOA  & 113.6570  & 26.26 & 0.363835  & 71.971  & 24.25 \\
    \bottomrule
  \end{tabular}
  }
\end{table}

Significant differences emerge in the ability of QUBO solutions generated by different methods to retain their structural quality in the simulation layer. The results show that the average total cost and average delay time of all methods increase relative to the normal scenario, but the magnitudes of these increases differ substantially.

Numerically, QGA-QAOA still performs best in the dynamic scenario, with an average total cost of 100.4085 and an average delay time of 62.602 min. QGA achieves an average total cost of 102.4755 and an average delay time of 64.453 min, while QPSO yields 105.6825 and 68.274 min, respectively. By contrast, the conventional simulated annealing algorithm rises to 154.1340 in average total cost and 95.014 min in average delay time. The average total-cost reduction rates of the quantum-inspired and hybrid methods range from 4.28\% to 26.26\%, and the average delay-reduction rates range from 4.37\% to 24.25\% relative to their conventional counterparts. These results show that operational fluctuations expose clear differences in the structural retention ability of decoded QUBO schemes.

This difference implies that performance in the dynamic scenario depends not only on the static departure order itself but also on whether the combined structure of departure sequence and track allocation can adapt to operational uncertainty. Solutions with lower average delay generally exhibit a more balanced departure rhythm in the origin section and more stable section connectivity, which helps suppress the accumulation of waiting time under local disturbances. Likewise, a lower average total cost suggests that delay is controlled without sharply increasing additional organizational costs such as track switching, same-track congestion, and platform occupation. Therefore, the dynamic experiment supports continued investigation of quantum-inspired and hybrid methods in disturbed dispatching environments within the proposed framework.
\FloatBarrier
\subsection{Analysis of Perturbation Robustness Results}
\begin{figure}[!htbp]
    \centering
    \includegraphics[width=1\linewidth]{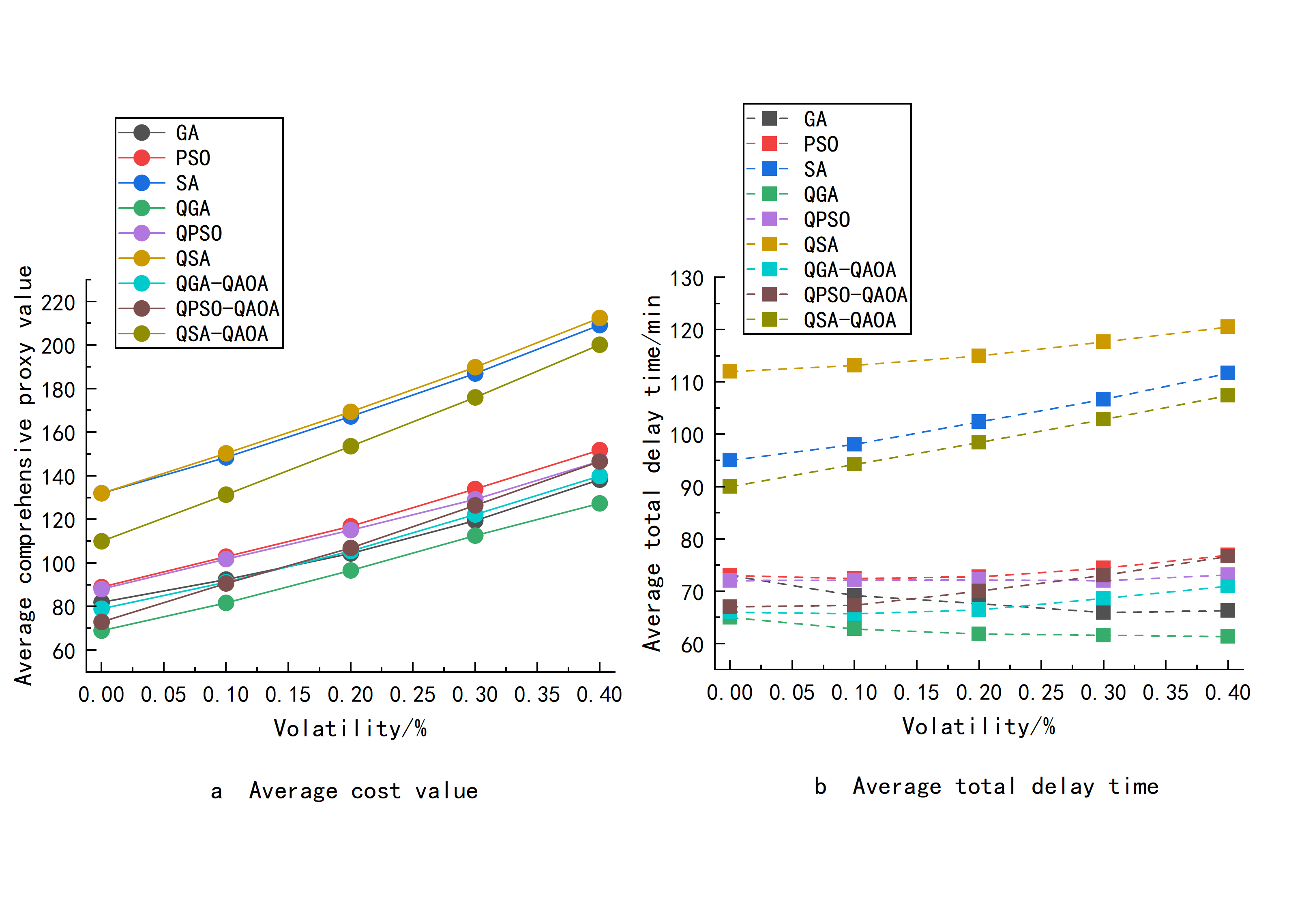}
    \caption{Robustness experiment results}
    \label{fig:Robustness experiment results}
\end{figure}

To further investigate the ability of the obtained schemes to withstand external fluctuations, the robustness experiment gradually increases the section fluctuation rate while repeatedly evaluating the operational performance of fixed combinatorial schemes. The focus here is no longer the quality of a single optimization run, but the extent to which different decoded schemes are prone to delay propagation, station occupation pressure, and rapid cost growth under intensified perturbations. The results are shown in Fig.~\ref{fig:Robustness experiment results}.
\FloatBarrier
Schemes generated by different algorithms exhibit different levels of tolerance to intensified perturbations, and these differences reflect the operational stability of the QUBO combinatorial structure. As the fluctuation rate increases, the total cost and delay level of all schemes rise accordingly, but the magnitudes of these increases are not consistent. This indicates that the combinatorial solutions obtained by different algorithms based on the QUBO model differ significantly in their sensitivity to operational perturbations.

From the perspective of total cost variation, when the fluctuation rate reaches 0.4, the average total cost of the QGA scheme is 127.38, whereas that of the conventional simulated annealing algorithm rises to 212.45, showing a substantially widened gap. This demonstrates that under intensified perturbations, the departure-order and track-allocation structures generated by different algorithms exhibit marked differences in their ability to suppress delay propagation and deterioration in resource occupation. Further comparison shows that the penalty related to siding occupancy is more sensitive to changes in the fluctuation rate. At a fluctuation rate of 0.4, the average siding penalty of the QGA scheme is only 2.2, whereas that of the conventional simulated annealing algorithm reaches 27.7. This indicates that operational perturbations in the current scenario tend to manifest first as dwell-time backlogs at intermediate stations and resource shortages within stations, and that schemes with better control over this process generally possess stronger structural stability.

These results indicate that the departure-order and track-allocation structure produced by the QUBO layer not only determines the cost level under the baseline scenario but also influences whether the decoded scheme is prone to station-resource congestion and delay propagation under intensified perturbations. The key implication of the robustness experiment is therefore not simply which method performs best, but how well the QUBO-generated scheme retains its structural quality when the operating environment deteriorates. Several quantum-inspired and hybrid methods exhibit comparatively strong robustness in this constructed setting, making them worthwhile targets for further methodological investigation.

\subsection{Analysis of Problem Scale Results}
\begin{figure}[!htbp]
    \centering
    \includegraphics[width=1\linewidth]{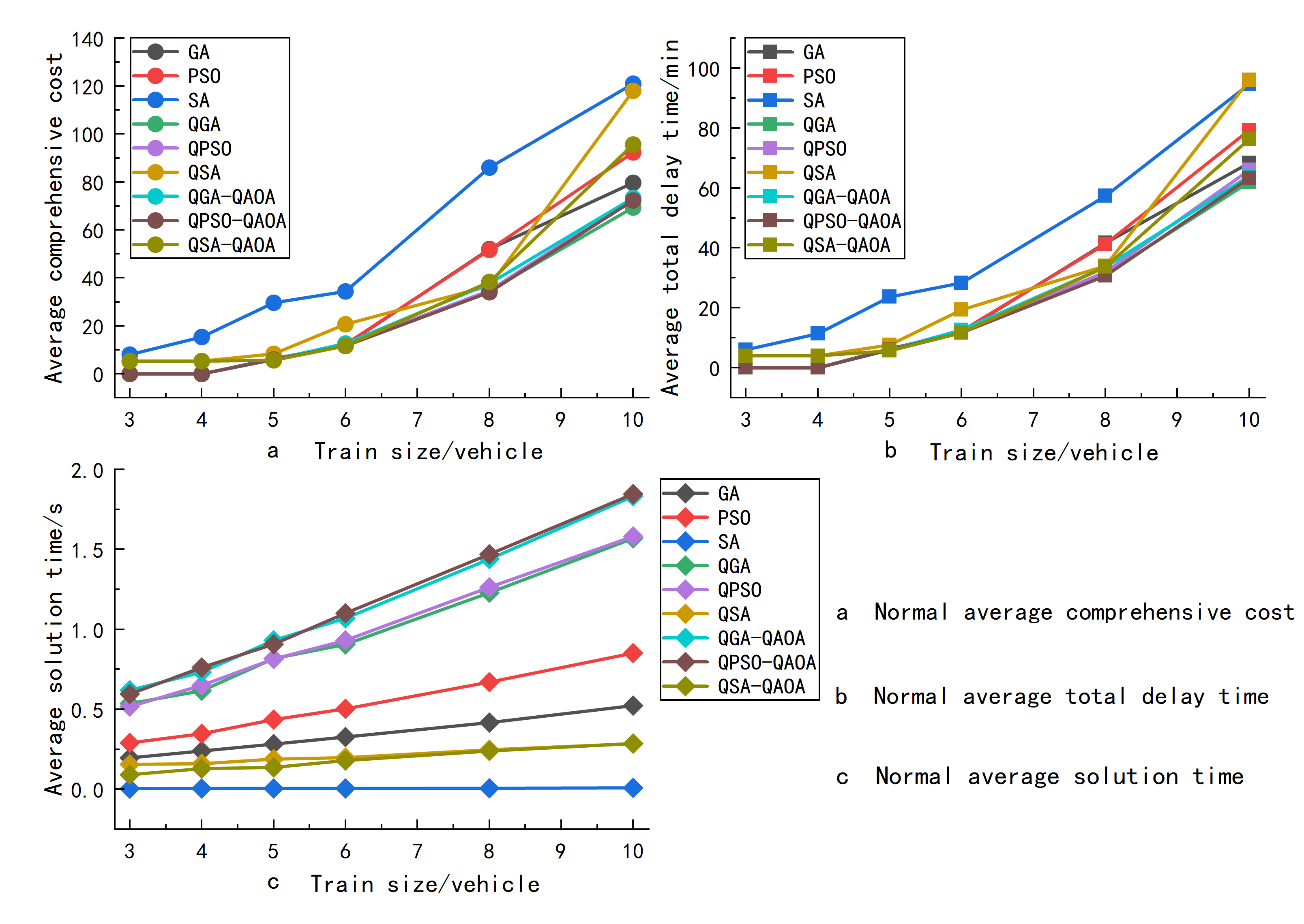}
    \caption{Experimental results of problem scale under normal conditions}
    \label{fig:Experimental results of problem scale under normal conditions}
\end{figure}

\begin{figure}[!htbp]
    \centering
    \includegraphics[width=1\linewidth]{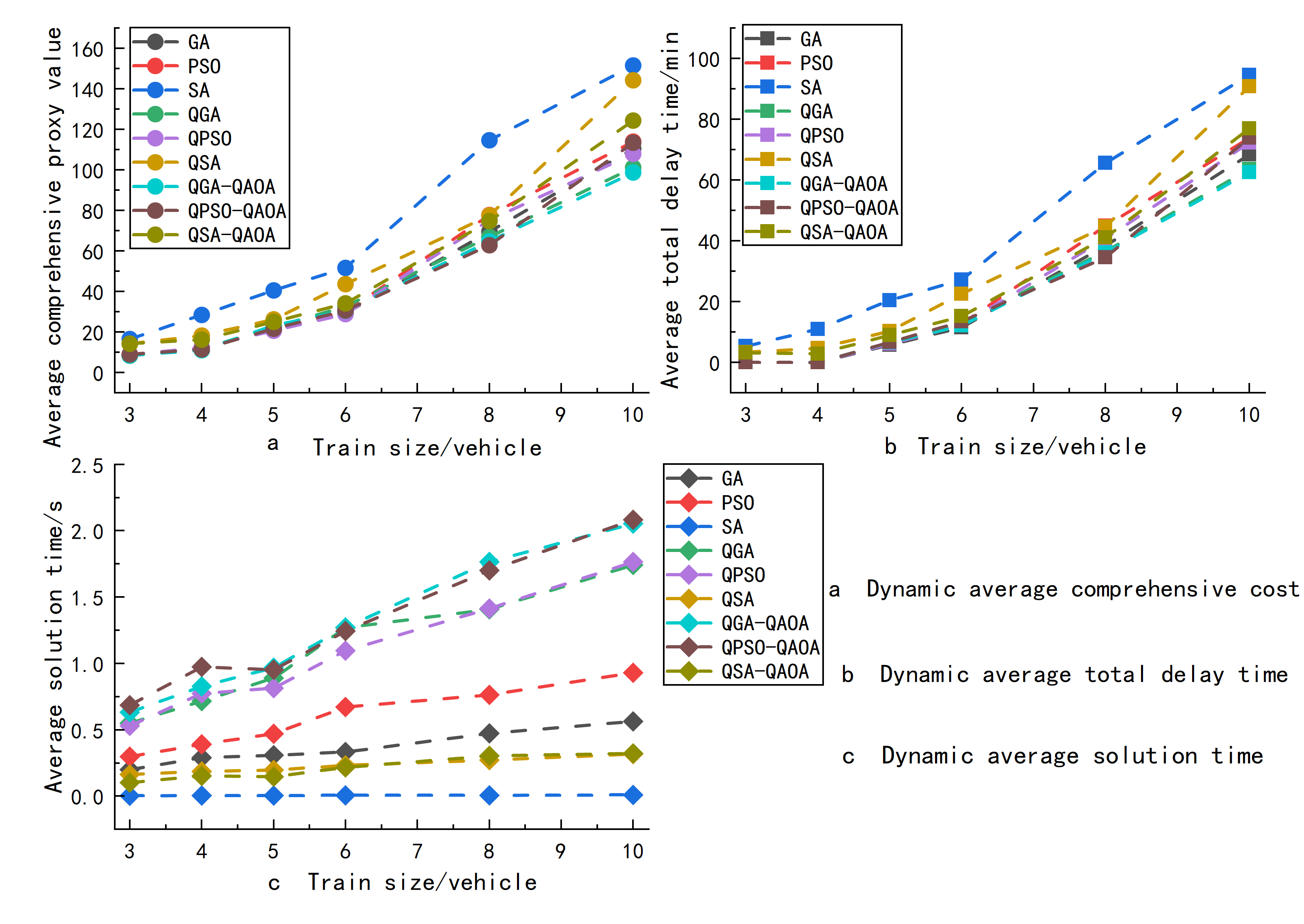}
    \caption{Experimental results of problem scale under dynamic conditions}
    \label{fig:Experimental results of problem scale under dynamic conditions}
\end{figure}

To investigate the applicability of different methods under intensified resource competition, scale-expansion experiments are carried out by gradually increasing the number of trains to be dispatched within a short period. As the number of trains increases, departure competition at the origin section, section-connection constraints, and dwell conflicts at intermediate stations intensify simultaneously. The resulting trends provide a controlled test of how different algorithms behave as the combinatorial decision space becomes more complex. The results under normal and dynamic conditions are shown in Fig.~\ref{fig:Experimental results of problem scale under normal conditions} and Fig.~\ref{fig:Experimental results of problem scale under dynamic conditions}, respectively.

The QUBO model continues to provide a unified representation for the concentrated departure scenario as the problem size increases, while the solution performance of different methods begins to diverge more clearly in the enlarged combinatorial space. The scale experiment shows that when the number of trains is small, the difference in average total cost across methods is limited, indicating that the QUBO formulation can represent the departure-order and track-allocation problem stably in small instances. As the number of trains increases, departure competition, section-connection constraints, and dwell conflicts intensify simultaneously, and the gap between methods becomes more pronounced.

In the normal scenario, the average total cost of all methods increases with problem scale, but the growth rates differ. QGA, QGA-QAOA, and QPSO show a relatively moderate increase under medium- and large-scale conditions, whereas the cost of conventional PSO and simulated annealing rises more rapidly. This indicates that the QUBO model remains comparable across scales while the adaptability differences among conventional, quantum-inspired, and hybrid methods become more visible in larger combinatorial spaces.

After introducing section-operation fluctuations, the scale effect becomes more pronounced. Under dynamic conditions, even small-scale scenarios no longer remain close to cost-free operation, and the total cost of medium- and large-scale scenarios increases further. This suggests that random deviations in running time are transmitted backward through section release and intermediate-station waiting, thereby amplifying the conflict accumulation caused by scale expansion. Several quantum-inspired and hybrid methods still maintain relatively low average total cost in medium- and large-scale dynamic scenarios, whereas some conventional methods become more vulnerable to concentrated waiting and station-resource shortage.

Overall, the QUBO formulation remains applicable after the train scale is expanded, indicating a certain degree of scalability for this class of concentrated departure problems. At the same time, the relative behavior of different methods becomes increasingly differentiated as complexity grows, suggesting that the medium- and large-scale applicability of quantum-inspired and hybrid methods merits further examination.
\section{Conclusions and Prospects}
This study investigates railway short-term concentrated departure scheduling and develops a layered optimization framework consisting of a QUBO-based combinatorial decision layer and an operation-evaluation layer. Under a unified modeling and simulation setting, conventional heuristics, quantum-inspired algorithms, and hybrid algorithms are compared systematically. The results show that departure-sequence assignment and section-track selection in short-term concentrated departure scenarios can be encoded within a unified QUBO formulation, and that the decoded solutions can be transformed into executable scheduling schemes whose operation diagrams, timetables, and statistical indicators are generated through simulation. This combined treatment of combinatorial optimization and operational-process verification provides a more complete means of characterizing sequence arrangement, resource coordination, and delay propagation in concentrated railway departure organization. Further analysis indicates that different solution methods exhibit clear performance differences under normal, disturbed, robustness, and scale-expansion settings. Several quantum-inspired and hybrid methods demonstrate good feasibility and methodological potential in the tested scenario, thereby providing a useful reference for future algorithm selection and scheme evaluation in railway dispatching studies.\par
However, the present case study still relies on a constructed test scenario and has not yet been validated using specific line data, station layouts, and original train diagrams from real operations. Future research can therefore extend the framework to actual railway corridors, more complex station structures, multidirectional coordination, and larger-scale network settings in order to examine the practical applicability of QUBO-based modeling and quantum-related algorithms in railway transportation organization more comprehensively.
\section*{Code Availability}
The source code and README files used in this study are publicly available at:
\url{https://github.com/yuncifor/Railway-Short-Term-Based-on-QUBO-and-Hybrid-Quantum-Algorithms}
\bibliographystyle{unsrt}
\bibliography{references}  

\end{document}